\documentclass[english,aps,twocolumn,prl]{revtex4}
\usepackage[T1]{fontenc}
\usepackage[latin9]{inputenc}
\usepackage{amsmath}
\usepackage{amssymb}
\usepackage{graphicx}

\makeatletter

\newcommand{\noun}[1]{\textsc{#1}}

\@ifundefined{textcolor}{}
{%
 \definecolor{BLACK}{gray}{0}
 \definecolor{WHITE}{gray}{1}
 \definecolor{RED}{rgb}{1,0,0}
 \definecolor{GREEN}{rgb}{0,1,0}
 \definecolor{BLUE}{rgb}{0,0,1}
 \definecolor{CYAN}{cmyk}{1,0,0,0}
 \definecolor{MAGENTA}{cmyk}{0,1,0,0}
 \definecolor{YELLOW}{cmyk}{0,0,1,0}
}

\makeatother

\usepackage{babel}
\begin{document}

\preprint{\pagebreak{}This line only printed with preprint option}

\title{Exact density functional obtained via the Levy constrained search}

\author{Paula Mori-Sánchez}

\affiliation{Departamento de Química and Instituto de Física de la Materia Condensada
(IFIMAC), Universidad Autónoma de Madrid, 28049, Madrid, Spain}

\author{Aron J. Cohen}

\affiliation{Max Planck Institute for Solid State Research, Heisenbergstrasse
1, 70569 Stuttgart, Germany}
\begin{abstract}
A stochastic minimization method for a real-space wavefunction, $\Psi({\bf r}_{1},{\bf r}_{2}\ldots{\bf r}_{n})$,
constrained to a chosen density, $\rho({\bf r})$, is developed. It
enables the explicit calculation of the Levy constrained search $F[\rho]=\min_{\Psi\rightarrow\rho}\langle\Psi|\hat{T}+\hat{V}_{ee}|\Psi\rangle$
(\emph{Proc. Natl. Acad. Sci. }\textbf{76} 6062 (1979)), that gives
the exact functional of density functional theory. This general method
is illustrated in the evaluation of $F[\rho]$ for two-electron densities
in one dimension with a soft-Coulomb interaction. Additionally, procedures
are given to determine the first and second functional derivatives,
$\frac{\delta F}{\delta\rho({\bf r})}$ and $\frac{\delta^{2}F}{\delta\rho({\bf r})\delta\rho({\bf r}')}$.
For a chosen external potential, $v({\bf r})$, the functional and
its derivatives are used in minimizations only over densities to give
the exact energy, $E_{v}$ without needing to solve the Schrödinger
equation. 
\end{abstract}
\maketitle

The electron density, $\rho({\bf r}),$ is the central object of density
functional theory (DFT). The foundational theorems \cite{Hohenberg64864,Levy796062}
prove the existence of one universal functional in the space of densities,
$F[\rho],$ that contains the necessary information to give the exact
many-body ground-state energy of all possible systems. In virtually
all calculations in the literature, the map is approximated within
the context of Kohn-Sham (KS) non-interacting reference \cite{Kohn651133}
with approximate functionals such as PBE and B3LYP \cite{Perdew963865,Becke935648}.
However, the exact map, $F[\rho]$, utilizes the exact many body-wavefunction
$\Psi({\bf r}_{1}\ldots{\bf r}_{N})$, where $\rho({\bf r})=N\int|\Psi({\bf r},{\bf r}_{2}\dots{\bf r}_{N})|^{2}{\rm d}{\bf r}_{2}\ldots{\rm d}{\bf r}_{N}$,
and does not involve KS. The Levy constrained search \cite{Levy796062}
is defined by considering only many-body wavefunctions that all integrate
to the same one-electron density, $\Psi\rightarrow\rho$, to define
the exact density functional
\begin{equation}
F[\rho]=\min_{\Psi\rightarrow\rho}\langle\Psi|\hat{T}+\hat{V}_{ee}|\Psi\rangle.\label{eq:Levy}
\end{equation}
For each given density, $\rho({\bf r})$, there is one minimizing
wavefunction, $\Psi_{\rho}^{{\rm min}}$, and one value of the functional,
$F[\rho]=\langle\Psi_{\rho}^{\min}|\hat{T}+\hat{V}_{ee}|\Psi_{\rho}^{\min}\rangle$.
The whole formalism is constructed to be completely independent of
any potential, $v({\bf r})$, or the more complicated questions whether
the $\rho({\bf r})$ is the ground state of a potential ($v$-representability).
However, this one universal functional can be used in a minimization
solely over the space of all possible densities to give the exact
ground-state many-body energy of the Schrödinger equation for every
external potential
\begin{equation}
E_{v}[\rho]={\rm \min_{\rho}}\left\{ F[\rho]+\int v({\bf r})\rho({\bf r}){\rm d}{\bf r}\right\} .\label{eq:LevyMin}
\end{equation}
These are the foundational equations of DFT, however they have not
been numerically realized. In this Letter, we develop a method to
explicitly carry out the constrained search only over many-body wavefunctions
that integrate to the same density, $\Psi\rightarrow\rho$ as in Eq.
(\ref{eq:Levy}) and, furthermore, to carry out the minimization over
densities in Eq. (\ref{eq:LevyMin}), as an alternative to solving
the Schrödinger equation. The importance of the exact functional is
demonstrated for strongly correlated systems. 

\begin{figure}[t]
\includegraphics[angle=-90,width=1\columnwidth]{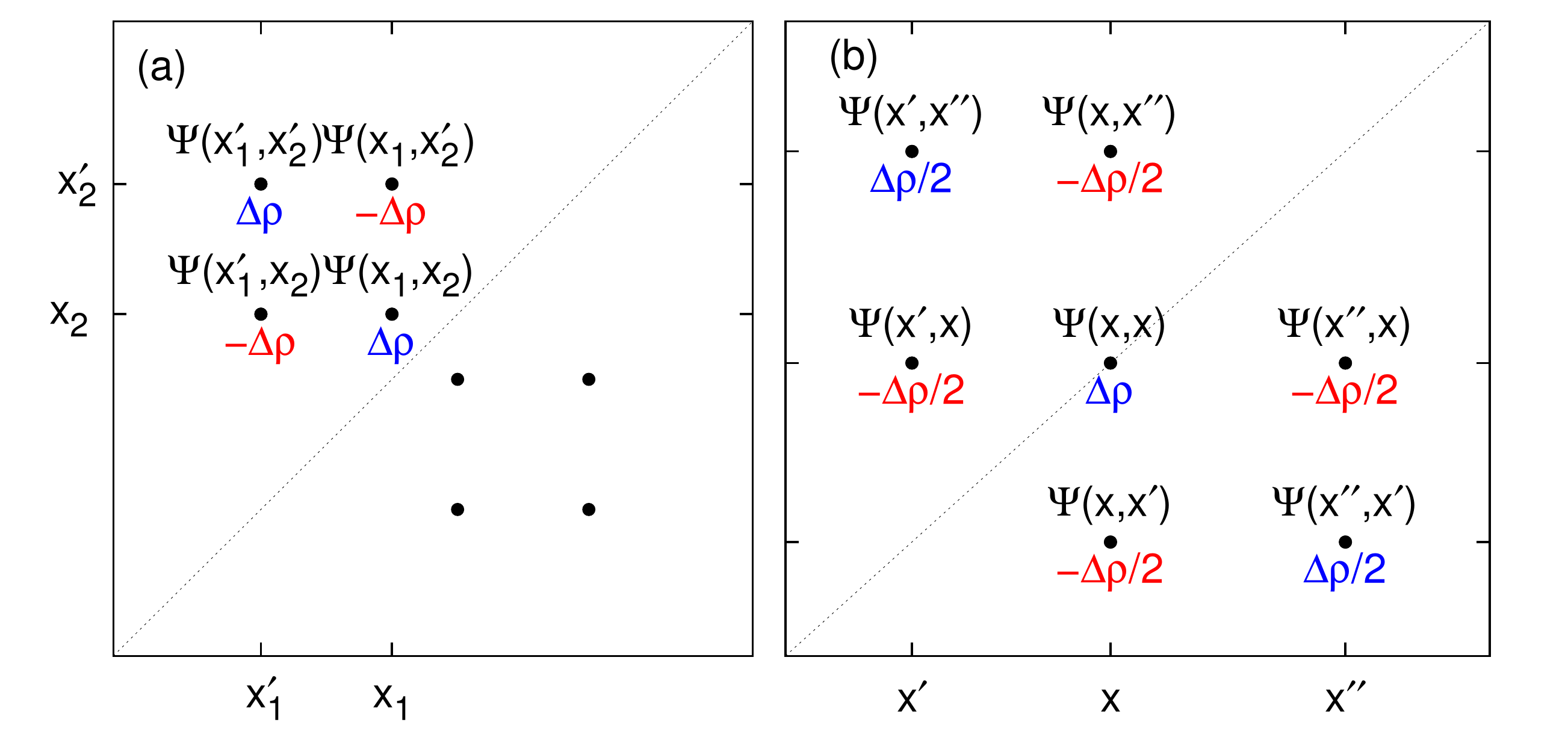}\caption{Example moves that maintain integration of a two-electron wavefunction
to a target density: (a) move for both singlet and triplet wavefunctions,
the bottom right would be moved symmetrically (anti-symmetrically)
for singlet (triplets); (b) move for the singlet wave function to
vary the diagonal elements. \label{fig:Example-Monte-Carlo-moves}}
\end{figure}

\begin{figure*}
\includegraphics[bb=0bp 200bp 792bp 612bp,width=1\textwidth]{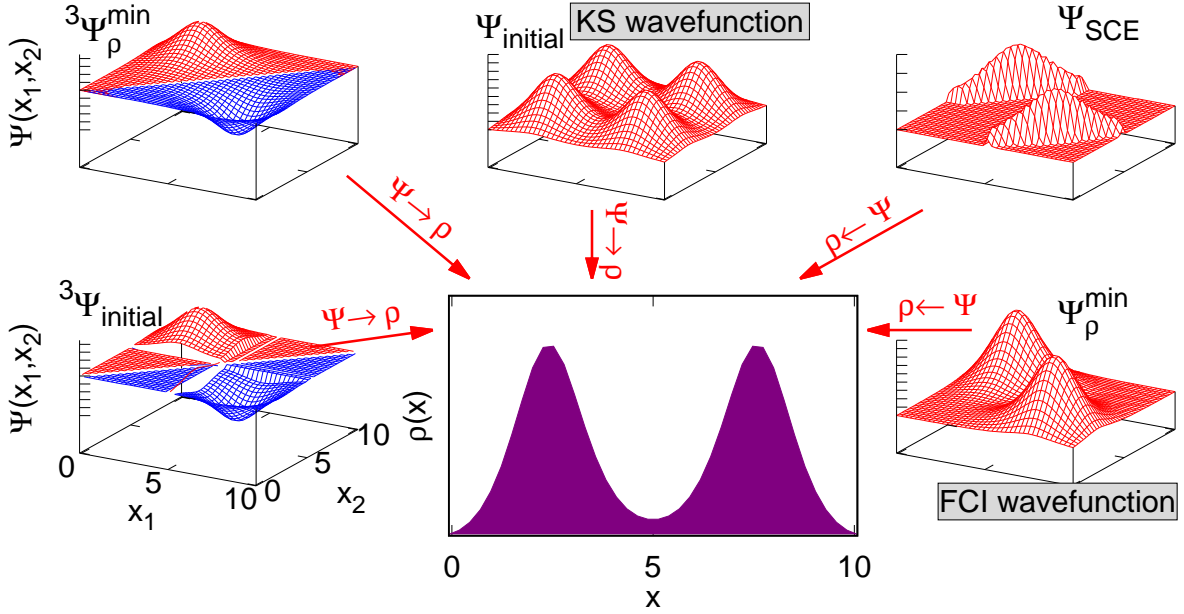}\caption{Different singlet and triplet wavefunctions, ranging from non-interacting
KS wavefunction, to fully interacting wavefunction and the wavefunction
for strictly correlated electrons (SCE) \cite{GoriGiorgi09166402},
that all yield the same one-dimensional density. \label{fig:Different-wavefunctions}}
\end{figure*}

To carry out the constrained search to a target density, denoted as
$\rho_{{\rm target}}$, we develop a stochastic procedure that contains
four key steps: 
\begin{description}
\item [{(1)}] Construct an initial wavefunction, $\Psi_{{\rm initial}}$,
that integrates to $\rho_{{\rm target}}$.
\item [{(2)}] Take a trial step in wavefunction space that maintains integration
to $\rho_{{\rm target}}$.
\item [{(3)}] Evaluate the many-body energy of this trial wavefunction.
\item [{(4)}] Accept or reject step and return to \textbf{\noun{(2)}} until
convergence. 
\end{description}
The fundamental aspect of this method is to define appropriate movements
in wavefunction space that keep the integration to the same density
in step (2). Consider an $N$-electron fermionic wavefunction that
can be separated into a spatial part, $\Psi({\bf r}_{1},{\bf r}_{2}\ldots{\bf r}_{N})$,
multiplied by a fixed spin part, $\sigma(s_{1},s_{2}\ldots s_{N})$.
A change in $\Psi({\bf R})$ at one point in Hilbert space ${\bf R}={\bf R}_{1},{\bf R}_{2}\ldots{\bf R}_{N}$
changes the density at $N$ separate points $\rho({\bf R}_{1}),\rho({\bf R}_{2})\ldots\rho({\bf R}_{N})$.
A second chosen point is needed, ${\bf R}^{\prime}={\bf R}_{1}^{\prime},{\bf R}_{2}^{\prime}\ldots{\bf R}_{N}^{\prime}$,
where ${\bf R}_{i}^{\prime}\ne{\bf R}_{1}\ {\rm or}\ {\bf R}_{2}\ {\rm or\ {\bf R}_{3}}\ldots{\rm or}\ {\bf R}_{N}\ \forall i$.
Next, define a replacement operator $\hat{P}_{i}$ which replaces
${\bf R}_{i}$ by ${\bf R}_{i}^{\prime}$, and take all possible replacements
of the original wavefunction, $\Psi_{l_{1}l_{2}\ldots l_{N}}=\Psi(\Pi_{i=1}^{N}\hat{P}_{i}^{l_{i}}{\bf R})$,
with $l_{i}=0\ {\rm or}\ 1$. Using this collections of wavefunctions
it is possible to construct a move that does not modify the density,
$\Psi_{l_{1}l_{2}\ldots l_{N}}^{2}\rightarrow\Psi_{l_{1}l_{2}\ldots l_{N}}^{2}+(-1)^{\sum_{i}^{N}l_{i}}\Delta$
for all binary numbers $l_{1}l_{2}\ldots l_{N}$ from $00\ldots0$
up to $11\ldots1$. The combined move changes $2^{N}$ points of the
original wavefunction (and all its symmetric or antisymmetric mirrored
parts) whilst maintaining the integration to $\rho_{{\rm target}}.$
This overall procedure illustrates the principle of many different
real-space wavefunctions, with different values of $\langle\Psi|\hat{T}+\hat{V}_{ee}|\Psi\rangle$,
that all yield identical densities, which is the key to performing
the Levy constrained search.

To demonstrate this method explicitly, we examine two-electron densities
in one-dimension, $\rho(x)$. The one-dimensional universe is described,
as in previous work \cite{Helbig11032503,WSBW12}, using a real space
grid with a softened Coulomb interaction, $\hat{V}_{ee}(x_{1},x_{2})=1/\sqrt{(x_{1}-x_{2})^{2}+1}$
and $\hat{T}=-\frac{1}{2}\frac{d^{2}}{dx^{2}}$. This work focuses
on two electron systems, where the overall wavefunction is antisymmetric
and separable, $\varPsi(x_{1}s_{1},x_{2}s_{2})=\Psi(x_{1},x_{2})\sigma(s_{1},s_{2})$,
with spin parts being the singlet or triplet, $\sigma(s_{1},s_{2})=\frac{1}{\sqrt{2}}(\alpha(s_{1})\beta(s_{2})\mp\alpha(s_{2})\beta(s_{1}))$.

Regarding the construction of $\Psi_{{\rm initial}}$ in step (1),
for the singlet state $\Psi_{{\rm initial}}(x_{1},x_{2})=\Psi_{{\rm KS}}=\sqrt{\rho(x_{1})\rho(x_{2}})/2$,
and for the triplet state a Gilbert construction \cite{Gilbert752111}
with a direct division of space gives two orbitals such that $^{3}\Psi_{{\rm initial}}(x_{1},x_{2})=[\phi_{1}(x_{1})\phi_{2}(x_{2})-\phi_{2}(x_{1})\phi_{1}(x_{2})]/\sqrt{2}$.
In this case, to define the appropriate movements of the wavefunction
in step (2) we consider ${\bf R}=x_{1},x_{2}$ and move the wavefunction
by a small amount $\epsilon_{x_{1}x_{2}}$, $\Psi^{{\rm trial}}(x_{1},x_{2})=\Psi(x_{1},x_{2})+\epsilon_{x_{1}x_{2}}$,
with a change in density 
\begin{equation}
\Delta\rho=\epsilon_{x_{1}x_{2}}^{2}+2\epsilon_{x_{1}x_{2}}\Psi(x_{1},x_{2})
\end{equation}
at two points, $\rho(x_{1})$ and $\rho(x_{2})$. So we randomly take
another point ${\bf R}^{\prime}=x_{1}^{\prime}$,$x_{2}^{\prime}\ne x_{1}$
or $x_{2}$ and consider moves of the wavefunction at $\Psi^{{\rm trial}}(x_{1},x_{2}^{\prime})=\Psi(x_{1},x_{2}^{\prime})+\epsilon_{x_{1}x_{2}^{\prime}}$
and $\Psi^{{\rm trial}}(x_{1}^{\prime},x_{2})=\Psi(x_{1}^{\prime},x_{2})+\epsilon_{x_{1}^{\prime}x_{2}}$
and $\Psi^{{\rm trial}}(x_{1}^{\prime},x_{2}^{\prime})=\Psi(x_{1}^{\prime},x_{2}^{\prime})+\epsilon_{x_{1}^{\prime}x_{2}^{\prime}}$.
The amount by which the wavefunction has to move at each of these
points is found by solving the quadratic equations 
\begin{equation}
\epsilon_{x_{i}x_{j}}^{2}+2\Psi(x_{i},x_{j})\epsilon_{x_{i}x_{j}}-\Delta\rho_{x_{i}x_{j}}=0\label{eq:quadratic}
\end{equation}
with $\Delta\rho_{x_{1}x_{2}^{\prime}}=\Delta\rho_{x_{1}^{\prime}x_{2}}=-\Delta\rho$
and $\Delta\rho_{x_{1}^{\prime}x_{2}^{\prime}}=\Delta\rho$. For the
case $x_{1}=x_{2}=x$ an alternative move is used with $\Delta\rho_{x,x'}=\Delta\rho_{x'x}=\Delta\rho_{xx''}=\Delta\rho_{x''x}=-\Delta\rho/2$
and $\Delta\rho_{x'x''}=\Delta\rho_{x''x'}=\Delta\rho/2$. These movements
are pictured in Fig. \ref{fig:Example-Monte-Carlo-moves}. If the
solution of any of the quadratic equations (Eq. (\ref{eq:quadratic}))
has no real roots, the whole move is rejected. Otherwise, the trial
energy, $E^{{\rm trial}}=\langle\Psi^{{\rm trial}}|\hat{T}+\hat{V}_{ee}|\Psi^{{\rm trial}}\rangle$,
is evaluated such that if it is lower than the current energy, the
step is accepted and otherwise it is rejected (steps (3) and (4)).
This is repeated many times to optimize the wavefunction. The best
scheme we have found initially uses a very small random move in the
wavefunction, $-10^{-10}\le\epsilon\le10^{-10}$, $|\epsilon|>10^{-14}$,
and if a step is successful a simple line-search is carried out by
multiplying the size of the step by 10, but with the same $x_{1},x_{2},x_{1}^{\prime}$
and $x_{2}^{\prime}$ (or $x,x'$ and $x''$), and this is repeated
until the move is rejected. 

This method allows us to carry out the constrained search efficiently
and to calculate the exact functional for any density. For example,
Fig. \ref{fig:Different-wavefunctions} shows a variety of wavefunctions
that yield the same target one-dimensional density. The minimizing
singlet and triplet wavefunctions ($\Psi_{\rho}^{{\rm min}}$ and
$^{3}\Psi_{\rho}^{{\rm min}})$ are found by carrying out the constrained
search as outlined above. Convergence is found in 5000 stochastic
cycles over the wavefunction, each cycle consisting of an attempted
move at each point of the wavefunction.\emph{ }

\begin{figure}
\includegraphics[bb=70bp 60bp 792bp 612bp,width=1\columnwidth]{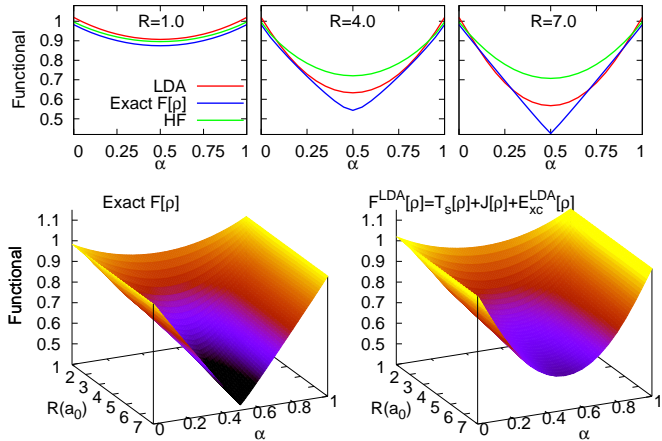}\caption{A cut of the landscape of the exact density functional for a set of
densities, $\rho(\alpha,R)=\alpha\rho_{{\rm He}}(0)+(1-\alpha)\rho_{{\rm He}}(R)\ (0\le\alpha\le1)$
compared with the LDA functional. $\alpha$ describes the transfer
of two electrons between two centers separated by a distance, $R$.
\label{fig:Functional Landscape} LDA breaks down in the strongly
correlated limit as $R$ is increased, especially for electron transfer.}
\end{figure}

Calculation of the Levy constrained search for one density gives one
point of $F[\rho]$, which in its entirety is a unique high-dimensional
surface in the $\Re$-dimensional space of densities. For example,
in a 100 grid point representation of $x$, the space of two-electron
densities, $\rho(x),$ would be 99 dimensional and $F[\rho(x)]$ can
be considered as a 99 dimensional function. Such a high dimensional
object cannot be visualized, but relevant cuts in restricted spaces
can be considered. For example, in Fig. \ref{fig:Functional Landscape},
the landscape of the exact functional is shown for a two-dimensional
slice of densities that are a convex linear combinations of two He
densities at a distance $R$, $\rho(\alpha,R)=\alpha\rho_{{\rm He}}(0)+(1-\alpha)\rho_{{\rm He}}(R)\ (0\le\alpha\le1)$,
where $\rho_{{\rm He}}(R)$ is a 1d-Helium atom density centered at
$R$, and $\alpha$ corresponds to the electron-transfer coordinate
between the two sites. $\alpha=0$ and $\alpha=1$ are Helium atoms
centered at different points and $\alpha=0.5$ corresponds to an H$_{2}$-like
density, whereas other values of $\alpha$ give fractional numbers
of electrons on the two centers. As the distance is increased, the
density becomes more strongly correlated, showing a very large difference
between the LDA and exact values of the functional ($\Delta F=0.139E_{h}=90\ {\rm kcal/mol}$,
for $R=7,\alpha=0.5)$ which is exactly the static correlation error
of systems like stretched H$_{2}$. However, much more important is
the global behavior on varying $\alpha$. At shorter distances, LDA
behaves similarly to the exact functional, but as the distance is
increased it exhibits a failure to describe the v-shape of the exact
functional with a qualitatively incorrect second derivative. It is
this failure in the $F^{{\rm LDA}}[\rho]$ surface that leads to the
incorrect description of charge transfer in strongly correlated systems
with DFT approximations \cite{Mori-Sanchez1414378}.

There has been other recent work on the exact functional via a Lieb
maximization \cite{Lieb83243}, based on a search over potentials
on which to carry out a many-body method such as full configuration
interaction (FCI) \cite{Colonna992828,coe_reverse_2009,Teale10164115,Wagner14045109,chen_pair_2015,entwistle_local_2016}
and in TDDFT \cite{nielsen_many-body_2013,ruggenthaler_existence_2015},
or Monte-Carlo based ideas to tackle the constrained search \cite{delle_site_electronic_2013,delle_site_levylieb_2015}.
However, the Lieb maximization would fail to converge if the density
is non-$v$-representable and the search over potentials becomes much
harder as the density becomes more strongly correlated \cite{wagner_guaranteed_2013,Wagner14045109}.
The complexity of all these methods often leads to conceptualizing
the functional $F[\rho]$ as an algorithm to give one number for a
chosen density. However, it is important to understand the functional
as a surface in the space of densities and one of the defining concepts
of a surface is differentiability. Thus, we now show a way to explicitly
calculate the exact functional derivatives, $\frac{\delta F[\rho]}{\delta\rho(x)}$
and $\frac{\delta^{2}F[\rho]}{\delta\rho(x)\delta\rho(x')}$. 

The first functional derivative can be determined from the converged
wavefunction, i.e. $\left.\frac{\partial F}{\partial\Psi}\right|_{\rho}=0$.
Thus, the differentiation with respect to $\Psi(x_{1},x_{2})$ gives
the Schrödinger-like equation:
\begin{equation}
\left(\hat{T}+\hat{V}_{ee}\right)\Psi(x_{1},x_{2})=\left(\frac{\delta F}{\delta\rho(x_{1})}+\frac{\delta F}{\delta\rho(x_{2})}\right)\Psi(x_{1},x_{2})
\end{equation}
Multiplying by $\Psi(x_{1},x_{2})$ and integrating (summing) over
$x_{2}$ gives a matrix equation
\begin{equation}
h_{x_{1}}=\sum_{x_{2}}M_{x_{1}x_{2}}d_{x_{2}}\label{eq:veq}
\end{equation}
where $d_{x_{2}}=\frac{\delta F}{\delta\rho(x_{2})}$ $M_{x_{1}x_{2}}=\Psi^{2}(x_{1},x_{2})+\frac{1}{2}\rho(x_{1})\delta_{x_{1}x_{2}}$
and $h_{x_{1}}=\int\Psi(x_{1},x_{2})\left(\hat{T}+\hat{V}_{ee}\right)\Psi(x_{1},x_{2}){\rm d}x_{2}$.
Solution of these simultaneous equations by inversion of the $M$
matrix gives the functional derivative. Another way to calculate the
first derivative is to make a set of changes $\left\{ \Delta_{i}\right\} $
to the density, defined by,
\begin{equation}
\Delta_{i}(x)=\left\{ \begin{array}{c}
\Delta_{i}(x_{i})=10^{-6}\\
\Delta_{i}(x_{j})=\frac{2-\rho(x_{i})-10^{-6}}{2-\rho(x_{i})}\rho(x_{j})\forall j\ne i
\end{array}\right.
\end{equation}
This is a small movement in the density at $x_{i}$ that is compensated
by moving all other points of the density the opposite way, such that
the overall normalization does not change, $\int\Delta_{i}(x){\rm d}x=0$.
The vectors of the finite difference approximation are formed
\begin{equation}
\lim_{\epsilon\rightarrow0}\left\{ \frac{F[\rho+\epsilon\Delta_{i}]-F[\rho]}{\epsilon}\right\} =F_{i}=\int\frac{\delta F[\rho]}{\delta\rho(x)}\Delta_{i}(x){\rm d}x.
\end{equation}
giving the set of equations 
\begin{equation}
F_{k}=\sum_{x}v_{x}\Delta_{kx},
\end{equation}
that is solved by construction of the pseudo-inverse, $\Delta_{kx}^{-1}$,
and applying it on the $F_{k}$ vector. The same machinery enables
the calculation of the second derivatives,
\begin{equation}
\lim_{\epsilon\rightarrow0}\left\{ \frac{\frac{\delta F[\rho+\epsilon\Delta_{i}]}{\delta\rho(x)}-\frac{\delta F[\rho]}{\delta\rho(x)}}{\epsilon}\right\} =\int\frac{\delta^{2}F[\rho]}{\delta\rho(x)\delta\rho(y)}\Delta_{i}(y){\rm d}y.
\end{equation}
By constructing the potential $v_{i}(x)$ corresponding to the density
$\rho+\Delta_{i}$, the second derivative Hessian matrix is built,
\begin{equation}
\frac{\delta^{2}F[\rho]}{\delta\rho(x)\delta\rho(y)}=\Delta_{ky}^{-1}[v_{k}(x)-v(x)].
\end{equation}
The Hessian matrix can be diagonalized, and if a negative eigenvalue
is found, the functional would be concave and it would correspond
to a non-$v$-representable density (see for example the SI of Ref
\cite{Cohen160142511}). 

\begin{figure}
\includegraphics[angle=-90,width=1\columnwidth]{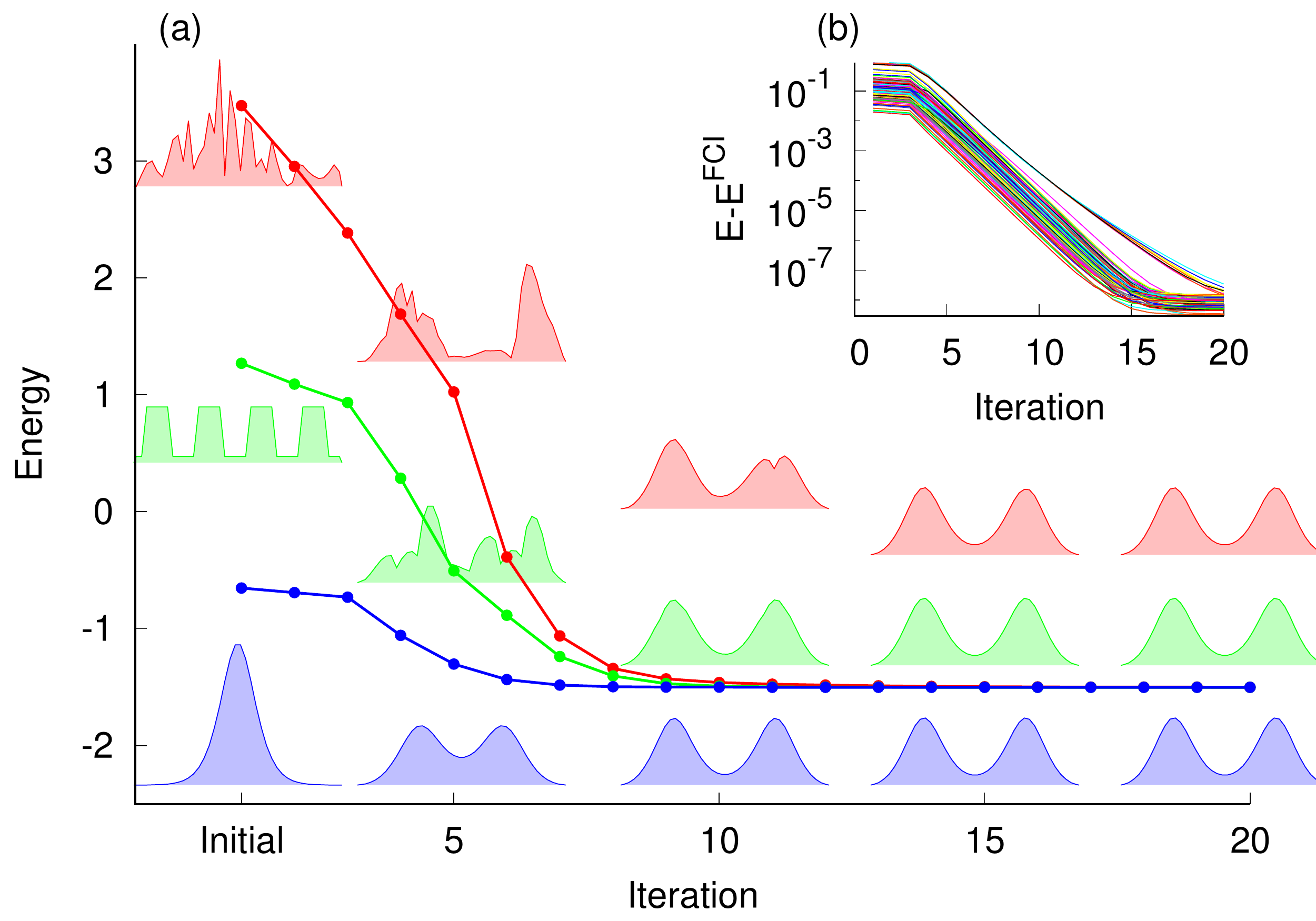}\caption{(a) Convergence of the energy and density for a one-dimensional H$_{2}$
potential, (FCI energy=$-1.500653$) starting from three different
initial densities: random (red), step-like (green) and He atom (blue).
Inset (b) shows the convergence of the energy to the FCI value for
100 different random molecular potentials, $v(x)=-Z/\sqrt{x^{2}+a}-Y/\sqrt{(x-R)^{2}+a}$.\label{fig:Optimisation}}
 
\end{figure}

The minimization of the energy in Eq. (\ref{eq:LevyMin}) can now
be carried out and treated as a general optimization problem using
the analytic first and second derivatives, $g_{y}=-\frac{\delta F[\rho]}{\delta\rho(y)}$
and $H_{xy}=\frac{\delta^{2}F[\rho]}{\delta\rho(x)\delta\rho(y)}$
. Therefore, for a given potential, $v_{y}=v(y)$, to best change
$\rho(x)$ to minimize the energy, a Newton step, ${\rm H}^{-1}{\rm g}$,
is taken
\begin{equation}
\Delta\rho(x)=m\left(\frac{\delta^{2}F}{\delta\rho(x)\delta\rho(y)}\right)^{-1}\left(-\frac{\delta F}{\delta\rho(y)}-v(y)\right)+C
\end{equation}
where the constant, $C$, enforces that the density is normalized,
$\int\Delta\rho(x){\rm d}x=0$. Usually $m=1$, though if the step
is not valid, i.e. $\rho(x)+\Delta\rho(x)\ngeq0\,\forall\,x$, a scaled
down step with $m=0.1$ is taken. This is only needed in the first
few steps of the optimization. None of the optimizations we have performed
have needed more than 35 steps and all of them have converged no matter
what the starting density is, as illustrated in Fig. \ref{fig:Optimisation}. 

In conclusion, we have explicitly carried out the Levy constrained
search using a stochastic optimization to give the exact functional
of DFT, $F[\rho]$. Millions of many-body wavefunctions are searched,
all integrating to the same $\rho({\bf r})$, and the minimizing wavefunction
that yields that $\rho({\bf r})$ and gives the lowest possible value
of $\langle\Psi_{\rho}|\hat{T}+\hat{V}_{ee}|\Psi_{\rho}\rangle$ is
found. Moreover, from this minimizing wavefunction, the first and
second functional derivatives are constructed and used in a direct
optimization of the total energy in density space for any external
potential. Thus, the exact many-body energy is obtained without solving
the Schrödinger equation and rapid convergence is seen from any starting
density, including strongly correlated systems. It is important to
view the exact functional as a calculable surface in density space,
as illustrated in Fig. \ref{fig:Functional Landscape}, highlighting
many avenues for future investigation, from the exact adiabatic connection
to the critical question of how to build more accurate approximations
to $F[\rho]$, for this system and beyond.

P.M.S. acknowledges funding from MINECO Grant No. FIS2015-64886-C5-5-P. 

\bibliographystyle{apsrev4-1}
%\bibliography{/Users/ajc54/Documents/allnewmodified}

\begin{thebibliography}{23}%
\makeatletter
\providecommand \@ifxundefined [1]{%
 \@ifx{#1\undefined}
}%
\providecommand \@ifnum [1]{%
 \ifnum #1\expandafter \@firstoftwo
 \else \expandafter \@secondoftwo
 \fi
}%
\providecommand \@ifx [1]{%
 \ifx #1\expandafter \@firstoftwo
 \else \expandafter \@secondoftwo
 \fi
}%
\providecommand \natexlab [1]{#1}%
\providecommand \enquote  [1]{``#1''}%
\providecommand \bibnamefont  [1]{#1}%
\providecommand \bibfnamefont [1]{#1}%
\providecommand \citenamefont [1]{#1}%
\providecommand \href@noop [0]{\@secondoftwo}%
\providecommand \href [0]{\begingroup \@sanitize@url \@href}%
\providecommand \@href[1]{\@@startlink{#1}\@@href}%
\providecommand \@@href[1]{\endgroup#1\@@endlink}%
\providecommand \@sanitize@url [0]{\catcode `\\12\catcode `\$12\catcode
  `\&12\catcode `\#12\catcode `\^12\catcode `\_12\catcode `\%12\relax}%
\providecommand \@@startlink[1]{}%
\providecommand \@@endlink[0]{}%
\providecommand \url  [0]{\begingroup\@sanitize@url \@url }%
\providecommand \@url [1]{\endgroup\@href {#1}{\urlprefix }}%
\providecommand \urlprefix  [0]{URL }%
\providecommand \Eprint [0]{\href }%
\providecommand \doibase [0]{http://dx.doi.org/}%
\providecommand \selectlanguage [0]{\@gobble}%
\providecommand \bibinfo  [0]{\@secondoftwo}%
\providecommand \bibfield  [0]{\@secondoftwo}%
\providecommand \translation [1]{[#1]}%
\providecommand \BibitemOpen [0]{}%
\providecommand \bibitemStop [0]{}%
\providecommand \bibitemNoStop [0]{.\EOS\space}%
\providecommand \EOS [0]{\spacefactor3000\relax}%
\providecommand \BibitemShut  [1]{\csname bibitem#1\endcsname}%
\let\auto@bib@innerbib\@empty
%</preamble>
\bibitem [{\citenamefont {Hohenberg}\ and\ \citenamefont
  {Kohn}(1964)}]{Hohenberg64864}%
  \BibitemOpen
  \bibfield  {author} {\bibinfo {author} {\bibfnamefont {P.}~\bibnamefont
  {Hohenberg}}\ and\ \bibinfo {author} {\bibfnamefont {W.}~\bibnamefont
  {Kohn}},\ }\href@noop {} {\bibfield  {journal} {\bibinfo  {journal} {Phys.
  Rev.}\ }\textbf {\bibinfo {volume} {136}},\ \bibinfo {pages} {B864} (\bibinfo
  {year} {1964})}\BibitemShut {NoStop}%
\bibitem [{\citenamefont {Levy}(1979)}]{Levy796062}%
  \BibitemOpen
  \bibfield  {author} {\bibinfo {author} {\bibfnamefont {M.}~\bibnamefont
  {Levy}},\ }\href@noop {} {\bibfield  {journal} {\bibinfo  {journal} {Proc.
  Natl. Acad. Sci. USA}\ }\textbf {\bibinfo {volume} {76}},\ \bibinfo {pages}
  {6062} (\bibinfo {year} {1979})}\BibitemShut {NoStop}%
\bibitem [{\citenamefont {Kohn}\ and\ \citenamefont {Sham}(1965)}]{Kohn651133}%
  \BibitemOpen
  \bibfield  {author} {\bibinfo {author} {\bibfnamefont {W.}~\bibnamefont
  {Kohn}}\ and\ \bibinfo {author} {\bibfnamefont {L.~J.}\ \bibnamefont
  {Sham}},\ }\href@noop {} {\bibfield  {journal} {\bibinfo  {journal} {Phys.
  Rev.}\ }\textbf {\bibinfo {volume} {140}},\ \bibinfo {pages} {A1133}
  (\bibinfo {year} {1965})}\BibitemShut {NoStop}%
\bibitem [{\citenamefont {Perdew}\ \emph {et~al.}(1996)\citenamefont {Perdew},
  \citenamefont {Burke},\ and\ \citenamefont {Ernzerhof}}]{Perdew963865}%
  \BibitemOpen
  \bibfield  {author} {\bibinfo {author} {\bibfnamefont {J.~P.}\ \bibnamefont
  {Perdew}}, \bibinfo {author} {\bibfnamefont {K.}~\bibnamefont {Burke}}, \
  and\ \bibinfo {author} {\bibfnamefont {M.}~\bibnamefont {Ernzerhof}},\
  }\href@noop {} {\bibfield  {journal} {\bibinfo  {journal} {Phys. Rev. Lett.}\
  }\textbf {\bibinfo {volume} {77}},\ \bibinfo {pages} {3865} (\bibinfo {year}
  {1996})}\BibitemShut {NoStop}%
\bibitem [{\citenamefont {Becke}(1993)}]{Becke935648}%
  \BibitemOpen
  \bibfield  {author} {\bibinfo {author} {\bibfnamefont {A.~D.}\ \bibnamefont
  {Becke}},\ }\href@noop {} {\bibfield  {journal} {\bibinfo  {journal} {J.
  Chem. Phys.}\ }\textbf {\bibinfo {volume} {98}},\ \bibinfo {pages} {5648}
  (\bibinfo {year} {1993})}\BibitemShut {NoStop}%
\bibitem [{\citenamefont {Gori-Giorgi}\ \emph {et~al.}(2009)\citenamefont
  {Gori-Giorgi}, \citenamefont {Seidl},\ and\ \citenamefont
  {Vignale}}]{GoriGiorgi09166402}%
  \BibitemOpen
  \bibfield  {author} {\bibinfo {author} {\bibfnamefont {P.}~\bibnamefont
  {Gori-Giorgi}}, \bibinfo {author} {\bibfnamefont {M.}~\bibnamefont {Seidl}},
  \ and\ \bibinfo {author} {\bibfnamefont {G.}~\bibnamefont {Vignale}},\
  }\href@noop {} {\bibfield  {journal} {\bibinfo  {journal} {Phys. Rev. Lett.}\
  }\textbf {\bibinfo {volume} {103}},\ \bibinfo {pages} {166402} (\bibinfo
  {year} {2009})}\BibitemShut {NoStop}%
\bibitem [{\citenamefont {Helbig}\ \emph {et~al.}(2011)\citenamefont {Helbig},
  \citenamefont {Fuks}, \citenamefont {Casula}, \citenamefont {Verstraete},
  \citenamefont {Marques}, \citenamefont {Tokatly},\ and\ \citenamefont
  {Rubio}}]{Helbig11032503}%
  \BibitemOpen
  \bibfield  {author} {\bibinfo {author} {\bibfnamefont {N.}~\bibnamefont
  {Helbig}}, \bibinfo {author} {\bibfnamefont {J.~I.}\ \bibnamefont {Fuks}},
  \bibinfo {author} {\bibfnamefont {M.}~\bibnamefont {Casula}}, \bibinfo
  {author} {\bibfnamefont {M.~J.}\ \bibnamefont {Verstraete}}, \bibinfo
  {author} {\bibfnamefont {M.~A.~L.}\ \bibnamefont {Marques}}, \bibinfo
  {author} {\bibfnamefont {I.~V.}\ \bibnamefont {Tokatly}}, \ and\ \bibinfo
  {author} {\bibfnamefont {A.}~\bibnamefont {Rubio}},\ }\href@noop {}
  {\bibfield  {journal} {\bibinfo  {journal} {Phys. Rev. A}\ }\textbf {\bibinfo
  {volume} {83}},\ \bibinfo {pages} {032503} (\bibinfo {year}
  {2011})}\BibitemShut {NoStop}%
\bibitem [{\citenamefont {Wagner}\ \emph {et~al.}(2012)\citenamefont {Wagner},
  \citenamefont {Stoudenmire}, \citenamefont {Burke},\ and\ \citenamefont
  {White}}]{WSBW12}%
  \BibitemOpen
  \bibfield  {author} {\bibinfo {author} {\bibfnamefont {L.~O.}\ \bibnamefont
  {Wagner}}, \bibinfo {author} {\bibfnamefont {E.}~\bibnamefont {Stoudenmire}},
  \bibinfo {author} {\bibfnamefont {K.}~\bibnamefont {Burke}}, \ and\ \bibinfo
  {author} {\bibfnamefont {S.~R.}\ \bibnamefont {White}},\ }\href@noop {}
  {\bibfield  {journal} {\bibinfo  {journal} {Phys. Chem. Chem. Phys.}\
  }\textbf {\bibinfo {volume} {14}},\ \bibinfo {pages} {8581 } (\bibinfo {year}
  {2012})}\BibitemShut {NoStop}%
\bibitem [{\citenamefont {Gilbert}(1975)}]{Gilbert752111}%
  \BibitemOpen
  \bibfield  {author} {\bibinfo {author} {\bibfnamefont {T.~L.}\ \bibnamefont
  {Gilbert}},\ }\href@noop {} {\bibfield  {journal} {\bibinfo  {journal} {Phys.
  Rev. B}\ }\textbf {\bibinfo {volume} {12}},\ \bibinfo {pages} {2111}
  (\bibinfo {year} {1975})}\BibitemShut {NoStop}%
\bibitem [{\citenamefont {Mori-S\'anchez}\ and\ \citenamefont
  {Cohen}(2014)}]{Mori-Sanchez1414378}%
  \BibitemOpen
  \bibfield  {author} {\bibinfo {author} {\bibfnamefont {P.}~\bibnamefont
  {Mori-S\'anchez}}\ and\ \bibinfo {author} {\bibfnamefont {A.~J.}\
  \bibnamefont {Cohen}},\ }\href@noop {} {\bibfield  {journal} {\bibinfo
  {journal} {Phys. Chem. Chem. Phys.}\ }\textbf {\bibinfo {volume} {16}},\
  \bibinfo {pages} {14378} (\bibinfo {year} {2014})}\BibitemShut {NoStop}%
\bibitem [{\citenamefont {Lieb}(1983)}]{Lieb83243}%
  \BibitemOpen
  \bibfield  {author} {\bibinfo {author} {\bibfnamefont {E.~H.}\ \bibnamefont
  {Lieb}},\ }\href@noop {} {\bibfield  {journal} {\bibinfo  {journal} {Int. J.
  Quant. Chem.}\ }\textbf {\bibinfo {volume} {24}},\ \bibinfo {pages} {243}
  (\bibinfo {year} {1983})}\BibitemShut {NoStop}%
\bibitem [{\citenamefont {Colonna}\ and\ \citenamefont
  {Savin}(1999)}]{Colonna992828}%
  \BibitemOpen
  \bibfield  {author} {\bibinfo {author} {\bibfnamefont {F.}~\bibnamefont
  {Colonna}}\ and\ \bibinfo {author} {\bibfnamefont {A.}~\bibnamefont
  {Savin}},\ }\href@noop {} {\bibfield  {journal} {\bibinfo  {journal} {J.
  Chem. Phys.}\ }\textbf {\bibinfo {volume} {110}},\ \bibinfo {pages} {2828}
  (\bibinfo {year} {1999})}\BibitemShut {NoStop}%
\bibitem [{\citenamefont {Coe}\ \emph {et~al.}(2009)\citenamefont {Coe},
  \citenamefont {Capelle},\ and\ \citenamefont {D'Amico}}]{coe_reverse_2009}%
  \BibitemOpen
  \bibfield  {author} {\bibinfo {author} {\bibfnamefont {J.~P.}\ \bibnamefont
  {Coe}}, \bibinfo {author} {\bibfnamefont {K.}~\bibnamefont {Capelle}}, \ and\
  \bibinfo {author} {\bibfnamefont {I.}~\bibnamefont {D'Amico}},\ }\href@noop
  {} {\bibfield  {journal} {\bibinfo  {journal} {Phys. Rev. A}\ }\textbf
  {\bibinfo {volume} {79}},\ \bibinfo {pages} {032504} (\bibinfo {year}
  {2009})}\BibitemShut {NoStop}%
\bibitem [{\citenamefont {Teale}\ \emph {et~al.}(2010)\citenamefont {Teale},
  \citenamefont {Coriani},\ and\ \citenamefont {Helgaker}}]{Teale10164115}%
  \BibitemOpen
  \bibfield  {author} {\bibinfo {author} {\bibfnamefont {A.~M.}\ \bibnamefont
  {Teale}}, \bibinfo {author} {\bibfnamefont {S.}~\bibnamefont {Coriani}}, \
  and\ \bibinfo {author} {\bibfnamefont {T.}~\bibnamefont {Helgaker}},\
  }\href@noop {} {\bibfield  {journal} {\bibinfo  {journal} {J. Chem. Phys.}\
  }\textbf {\bibinfo {volume} {132}},\ \bibinfo {pages} {164115} (\bibinfo
  {year} {2010})}\BibitemShut {NoStop}%
\bibitem [{\citenamefont {Wagner}\ \emph {et~al.}(2014)\citenamefont {Wagner},
  \citenamefont {Baker}, \citenamefont {Stoudenmire}, \citenamefont {Burke},\
  and\ \citenamefont {White}}]{Wagner14045109}%
  \BibitemOpen
  \bibfield  {author} {\bibinfo {author} {\bibfnamefont {L.~O.}\ \bibnamefont
  {Wagner}}, \bibinfo {author} {\bibfnamefont {T.~E.}\ \bibnamefont {Baker}},
  \bibinfo {author} {\bibfnamefont {E.~M.}\ \bibnamefont {Stoudenmire}},
  \bibinfo {author} {\bibfnamefont {K.}~\bibnamefont {Burke}}, \ and\ \bibinfo
  {author} {\bibfnamefont {S.~R.}\ \bibnamefont {White}},\ }\href@noop {}
  {\bibfield  {journal} {\bibinfo  {journal} {Phys. Rev. B}\ }\textbf {\bibinfo
  {volume} {90}},\ \bibinfo {pages} {045109} (\bibinfo {year}
  {2014})}\BibitemShut {NoStop}%
\bibitem [{\citenamefont {Chen}\ and\ \citenamefont
  {Friesecke}(2015)}]{chen_pair_2015}%
  \BibitemOpen
  \bibfield  {author} {\bibinfo {author} {\bibfnamefont {H.}~\bibnamefont
  {Chen}}\ and\ \bibinfo {author} {\bibfnamefont {G.}~\bibnamefont
  {Friesecke}},\ }\href@noop {} {\bibfield  {journal} {\bibinfo  {journal}
  {Multiscale Model. Simul.}\ }\textbf {\bibinfo {volume} {13}},\ \bibinfo
  {pages} {1259} (\bibinfo {year} {2015})}\BibitemShut {NoStop}%
\bibitem [{\citenamefont {Entwistle}\ \emph {et~al.}(2016)\citenamefont
  {Entwistle}, \citenamefont {Hodgson}, \citenamefont {Wetherell},
  \citenamefont {Longstaff}, \citenamefont {Ramsden},\ and\ \citenamefont
  {Godby}}]{entwistle_local_2016}%
  \BibitemOpen
  \bibfield  {author} {\bibinfo {author} {\bibfnamefont {M.~T.}\ \bibnamefont
  {Entwistle}}, \bibinfo {author} {\bibfnamefont {M.~J.~P.}\ \bibnamefont
  {Hodgson}}, \bibinfo {author} {\bibfnamefont {J.}~\bibnamefont {Wetherell}},
  \bibinfo {author} {\bibfnamefont {B.}~\bibnamefont {Longstaff}}, \bibinfo
  {author} {\bibfnamefont {J.~D.}\ \bibnamefont {Ramsden}}, \ and\ \bibinfo
  {author} {\bibfnamefont {R.~W.}\ \bibnamefont {Godby}},\ }\href@noop {}
  {\bibfield  {journal} {\bibinfo  {journal} {Phys. Rev. B}\ }\textbf {\bibinfo
  {volume} {94}} (\bibinfo {year} {2016})}\BibitemShut {NoStop}%
\bibitem [{\citenamefont {Nielsen}\ \emph {et~al.}(2013)\citenamefont
  {Nielsen}, \citenamefont {Ruggenthaler},\ and\ \citenamefont {van
  Leeuwen}}]{nielsen_many-body_2013}%
  \BibitemOpen
  \bibfield  {author} {\bibinfo {author} {\bibfnamefont {S.~E.~B.}\
  \bibnamefont {Nielsen}}, \bibinfo {author} {\bibfnamefont {M.}~\bibnamefont
  {Ruggenthaler}}, \ and\ \bibinfo {author} {\bibfnamefont {R.}~\bibnamefont
  {van Leeuwen}},\ }\href@noop {} {\bibfield  {journal} {\bibinfo  {journal}
  {(Europhys. Lett.)}\ }\textbf {\bibinfo {volume} {101}},\ \bibinfo {pages}
  {33001} (\bibinfo {year} {2013})}\BibitemShut {NoStop}%
\bibitem [{\citenamefont {Ruggenthaler}\ \emph {et~al.}(2015)\citenamefont
  {Ruggenthaler}, \citenamefont {Penz},\ and\ \citenamefont {van
  Leeuwen}}]{ruggenthaler_existence_2015}%
  \BibitemOpen
  \bibfield  {author} {\bibinfo {author} {\bibfnamefont {M.}~\bibnamefont
  {Ruggenthaler}}, \bibinfo {author} {\bibfnamefont {M.}~\bibnamefont {Penz}},
  \ and\ \bibinfo {author} {\bibfnamefont {R.}~\bibnamefont {van Leeuwen}},\
  }\href@noop {} {\bibfield  {journal} {\bibinfo  {journal} {J. Phys. Cond.
  Matt.}\ }\textbf {\bibinfo {volume} {27}},\ \bibinfo {pages} {203202}
  (\bibinfo {year} {2015})}\BibitemShut {NoStop}%
\bibitem [{\citenamefont {Delle~Site}\ \emph {et~al.}(2013)\citenamefont
  {Delle~Site}, \citenamefont {Ghiringhelli},\ and\ \citenamefont
  {Ceperley}}]{delle_site_electronic_2013}%
  \BibitemOpen
  \bibfield  {author} {\bibinfo {author} {\bibfnamefont {L.}~\bibnamefont
  {Delle~Site}}, \bibinfo {author} {\bibfnamefont {L.~M.}\ \bibnamefont
  {Ghiringhelli}}, \ and\ \bibinfo {author} {\bibfnamefont {D.~M.}\
  \bibnamefont {Ceperley}},\ }\href@noop {} {\bibfield  {journal} {\bibinfo
  {journal} {Int. J. Quantum Chem.}\ }\textbf {\bibinfo {volume} {113}},\
  \bibinfo {pages} {155} (\bibinfo {year} {2013})}\BibitemShut {NoStop}%
\bibitem [{\citenamefont {Delle~Site}(2015)}]{delle_site_levylieb_2015}%
  \BibitemOpen
  \bibfield  {author} {\bibinfo {author} {\bibfnamefont {L.}~\bibnamefont
  {Delle~Site}},\ }\href@noop {} {\bibfield  {journal} {\bibinfo  {journal}
  {Chem. Phys. Lett.}\ }\textbf {\bibinfo {volume} {619}},\ \bibinfo {pages}
  {148} (\bibinfo {year} {2015})}\BibitemShut {NoStop}%
\bibitem [{\citenamefont {Wagner}\ \emph {et~al.}(2013)\citenamefont {Wagner},
  \citenamefont {Stoudenmire}, \citenamefont {Burke},\ and\ \citenamefont
  {White}}]{wagner_guaranteed_2013}%
  \BibitemOpen
  \bibfield  {author} {\bibinfo {author} {\bibfnamefont {L.~O.}\ \bibnamefont
  {Wagner}}, \bibinfo {author} {\bibfnamefont {E.~M.}\ \bibnamefont
  {Stoudenmire}}, \bibinfo {author} {\bibfnamefont {K.}~\bibnamefont {Burke}},
  \ and\ \bibinfo {author} {\bibfnamefont {S.~R.}\ \bibnamefont {White}},\
  }\href@noop {} {\bibfield  {journal} {\bibinfo  {journal} {Phys. Rev. Lett.}\
  }\textbf {\bibinfo {volume} {111}} (\bibinfo {year} {2013})}\BibitemShut
  {NoStop}%
\bibitem [{\citenamefont {Cohen}\ and\ \citenamefont
  {Mori-Sánchez}(2016)}]{Cohen160142511}%
  \BibitemOpen
  \bibfield  {author} {\bibinfo {author} {\bibfnamefont {A.~J.}\ \bibnamefont
  {Cohen}}\ and\ \bibinfo {author} {\bibfnamefont {P.}~\bibnamefont
  {Mori-Sánchez}},\ }\href@noop {} {\bibfield  {journal} {\bibinfo  {journal}
  {Phys. Rev. A}\ }\textbf {\bibinfo {volume} {93}},\ \bibinfo {pages} {042511}
  (\bibinfo {year} {2016})}\BibitemShut {NoStop}%
\end{thebibliography}
%merlin.mbs apsrev4-1.bst 2010-07-25 4.21a (PWD, AO, DPC) hacked
%Control: key (0)
%Control: author (72) initials jnrlst
%Control: editor formatted (1) identically to author
%Control: production of article title (-1) disabled
%Control: page (0) single
%Control: year (1) truncated
%Control: production of eprint (0) enabled
%

\end{document}